\documentclass[12pt,a4paper]{article}%
\usepackage{epsfig}
\usepackage{amsmath}
\usepackage{amsfonts}
\usepackage{amssymb}
\usepackage{graphicx}
\usepackage[T1]{fontenc}
\usepackage[utf8]{inputenc}
\usepackage{authblk}%
\providecommand{\U}[1]{\protect\rule{.1in}{.1in}}
\begin{document}

\title{Non-Unitary evolution of quantum time-dependent non-Hermitian systems}
\author{Mustapha Maamache\thanks{E-mail: maamache@univ-setif.dz }\\Laboratoire de Physique Quantique et Syst\`{e}mes Dynamiques,\\Facult\'{e} des Sciences, Universit\'{e} Ferhat Abbas S\'{e}tif 1, S\'{e}tif
19000, Algeria. }
\date{}
\maketitle

\begin{abstract}
We provide a new perspective on non-Hermitian evolution in quantum mechanics
by emphasizing the same method as in the Hermitian quantum evolution. We first
give a precise description of the non unitary evolution, and collecting the
basic results around it and postulating the norm preserving. This cautionary
postulate imposing that the time evolution of a non Hermitian quantum system
preserves the inner products between the associated states must not be read
naively.\ We also give an example showing that the solutions of time-dependent
non Hermitian Hamiltonian systems given by a linear combination of SU(1,1) and
SU(2) are obtained thanks to time-dependent non-unitary transformation.



\end{abstract}

\section{Introduction}

\bigskip One of the postulates of quantum mechanics is that the Hamiltonian is
Hermitian, as this guarantees that the eigenvalues are real. This postulate
result from a set of postulates representing the minimal assumptions needed to
develop the theory of quantum mechanics. One of these postulates concerns the
time evolution of the state vector $\left\vert \psi(t)\right\rangle $ governed
by the Schr\"{o}dinger equation which describe how a state changes with time:%

\begin{equation}
i\hbar\frac{\partial}{\partial t}\left\vert \psi(t)\right\rangle =H\left\vert
\psi(t)\right\rangle \label{0}%
\end{equation}
where $H$ is the Hamiltonian operator corresponding to the total energy of the
system. The time dependent Schr\"{o}dinger equation is the most general way of
describing how a state changes with time.

The time evolution of the state of a quantum system described by $\left\vert
\psi(t)\right\rangle =U(t,t_{0})$ $\left\vert \psi(t_{0})\right\rangle $
preserves the normalization of the associated ket for some unitary operator
$U$ . The Hermiticity of the Hamiltonian $H$ guarantees that the energy
spectrum is real and the time evolution of the system is unitary. A direct
consequence of the hermiticity of $H$ is that the norm $\langle\psi
(t)\left\vert \psi(t)\right\rangle $ is time independent.

For explicitly time-dependent systems, solving the time-dependent
Schr\"{o}dinger equation is in general very difficult and it is very rare to
be able to find an exact solution. Various methods have been used to obtain
approximate solutions for such time-dependent problems. The usual methods are
the adiabatic approximation, the sudden approximation, and time-dependent
perturbation techniques. The existence of invariants \cite{Lewis1,Lewis2}
constants of the motion or first integral is one of central importance in the
study of such a system, be it classical or quantum. In the quantum case, Lewis
and Riesenfeld \cite{Lewis} first exploited the invariant operators to solve
quantum-mechanical problems. In particular, they have derived a simple
relation between eigenstates of invariants and solutions of the time-dependent
Schr\"{o}dinger equation and have applied it to the case of a quantal
oscillator with time-dependent frequency. Two models have been studied
extensively in the literature by several authors, for instance
\cite{1,2,3,4,5,6,mizrahi,7,8,9,10}. One of them is the time-dependent
generalized harmonic oscillator with the symmetry of the SU(1,1) dynamical
group, the other is the two-level system possessing an SU(2) symmetry. With
the help of the appropriate time-dependent unitary transformation instead of
the invariant operator, the solutions of SU(1, 1) and SU(2) time-dependent
quantum systems as well as the time-evolution operator are obtained in
\cite{mus1}. Time-dependent Hamiltonian systems are also of importance in
quantum optics.

The quantum mechanics is capable of working for some non-Hermitian quantum
systems. However, the Hermiticity is relaxed to be pseudo-Hermiticity
\cite{Scholz} or PT symmetry in non-Hermitian quantum mechanics, where is a
linear Hermitian or an anti-linear anti-Hermitian operator, and P and T stand
for the parity and time-reversal operators, respectively. The theories of
non-Hermitian quantum mechanics have been developed quickly in recent decades,
the reader can consulte the articles \cite{Carl1,most} and references cited therein.

Systems with time-dependent non-Hermitian Hamiltonian operators have been
studied in
\cite{Faria1,Faria2,most5,znojil2,znojil3,Bila,wang1,wang2,mus,fring1,fring2,khant,frith,luiz1,luiz2,mus2,mus3}%
. The most recent monograph \cite{bagarello} can be consulted for introduction
of the non-stationary theory.

Thus, we adapte the method based on a time-dependent unitary transformation of
time-dependent Hermitian Hamiltonians \cite{mus1} to solve the Schr\"{o}dinger
equation for the time-dependent non-Hermitian Hamiltonian systems given by a
linear combination of SU(1, 1) and SU(2) generators.

Because of the existence of a pseudo invariant operator, an SU(1, 1) and SU(2)
time-dependent non Hermitian quantum system must be integrable. For an
integrable system, the Hamiltonian can be transformed into a sum of
time-independent commuting operators through a time-dependent non-unitary
transformation. For this we introduce, in section 2, a formalism based on the
time-dependent non-unitary transformations and we show that the time-dependent
non-Hermitian Hamiltonian is related to an associated time-independent
Hamiltonian multiplied by an overall time-dependent factor. In section 3, we
illustrate our formalism introduced in the previous section by treating a
non-Hermitian SU(1, 1) and SU(2) time-dependent quantum problem and finding
the exact solution of the Schr\"{o}dinger equation without making recourse to
the pseudo- invariant operator theory as been done in \cite{mus2,mus3} or to
the technique presented in \cite{fring1,fring2}. Section 4, concludes our work.

\section{Formalism}

Consider the time-dependent Schr\"{o}dinger equation%

\begin{equation}
i\frac{\partial}{\partial t}\left\vert \psi(t)\right\rangle =H(t)\left\vert
\psi(t)\right\rangle , \label{1}%
\end{equation}
with $\hbar=1$ and $H(t)$ is the time-dependent non-Hermitian Hamiltonian
operator. Suppose that is a time-dependent non unitary transformation $V(t)$,%

\begin{equation}
\left\vert \phi(t)\right\rangle =V(t)\left\vert \psi(t)\right\rangle \label{2}%
\end{equation}
transform the state to $\left\vert \phi(t)\right\rangle $, which obeys the
Schr\"{o}dinger equation with the accordingly transformed non-Hermitian
Hamiltonian $\mathcal{H}(t)$%

\begin{equation}
i\frac{\partial}{\partial t}\left\vert \phi(t)\right\rangle =\mathcal{H}%
(t)\left\vert \phi(t)\right\rangle . \label{3}%
\end{equation}
In the above equation, the new Hamiltonian operator,
\begin{equation}
\mathcal{H}(t)=V(t)H(t)V^{-1}(t)+i\frac{\partial V(t)}{\partial t}V^{-1}(t),
\label{4}%
\end{equation}
is non-Hermitian. With the help of the appropriate non- unitary transformation
$V(t)$, we require that the solution of Schr\"{o}dinger equation looks like a
product of a state satisfying a stationary equation times a time-dependent
function. For this, we may make the transformed Hamiltonian $\mathcal{H}(t)$
possessing the following form:%

\begin{equation}
\mathcal{H}(t)=g(t)H_{0}, \label{5}%
\end{equation}
where the operator $H_{0}$ is time independent and $g(t)$ is a function of
time. The implication of these results is clear. The original time-dependent
non-Hermitian quantum problem posed through the Hamiltonian $H(t)$ and related
to an associated time independent Hamiltonian multiplied by an overall
time-dependent factor is completely solved. To get the form of the solution,
let $\left\vert \zeta_{n}\right\rangle $ be the eigenstate of $H_{0}$ with the
eigenvalue $\lambda_{n}$ i.e.%
\begin{equation}
H_{0}\left\vert \zeta_{n}\right\rangle =\lambda_{n}\left\vert \zeta
_{n}\right\rangle ,\text{ \ \ \ \ \ \ \ \ \ \ }\lambda_{n}=cte. \label{6}%
\end{equation}
Then, suppose that the system described by the Schr\"{o}dinger equation
(\ref{3}) is in the nth eigenstate of $H_{0}$ initially, at time $t$ it
evolves to the state
\begin{equation}
\left\vert \phi_{n}(t)\right\rangle =\exp\left(  i\lambda_{n}%
{\displaystyle\int\limits_{0}^{t}}
g(t^{\prime})dt^{\prime}\right)  \left\vert \zeta_{n}\right\rangle . \label{7}%
\end{equation}

Of note it follows immediately that the time evolution of a quantum system
described by $\left\vert \phi_{n}(t)\right\rangle $ doesn't preserve the
normalization i.e., the inner product of evolved states $\left\vert \phi
_{n}(t)\right\rangle $ depend on time:
\begin{equation}
\langle\phi_{n}(t)\left\vert \phi_{n}(t)\right\rangle =\exp\left(
\operatorname{Im}\left\{  \lambda_{n}%
{\displaystyle\int\limits_{0}^{t}}
g(t^{\prime}))dt^{\prime}\right\}  \right)  \langle\phi_{n}(0)\left\vert
\phi_{n}(0)\right\rangle \label{8}%
\end{equation}
At this stage, we will postulate, like the Hermitian case, that the time
evolution of a quantum system preserves, not just the normalization of the
quantum states, but also the inner products between the associated states
$\langle\phi_{n}(t)\left\vert \phi_{n}(t)\right\rangle =\langle\phi
_{n}(0)\left\vert \phi_{n}(0)\right\rangle $, implying that $g(t)$ and
$\lambda_{n}$ are reals and consequently the Hamiltonian $\mathcal{H}(t)$
should be Hermitian.

\bigskip As we are only changing our description of the system by changing
basis, we must preserve the inner product between vectors. Explicitly, from
preserving of this inner product between states $\left\vert \phi
_{n}(t)\right\rangle $, we can now define the inner product between states
$\left\vert \psi_{n}(t)\right\rangle =V^{-1}(t)\left\vert \phi_{n}%
(t)\right\rangle $ as%
\begin{equation}
\left\langle \psi_{n}(t)\right\vert V^{+}(t)V(t)\text{\ }\left\vert \psi
_{n}(t)\right\rangle =\left\langle \psi_{n}(0)\right\vert V^{+}%
(0)V(0)\text{\ }\left\vert \psi_{n}(0)\right\rangle \label{8'}%
\end{equation}
which has both a positive definite signature and leaves the norms of vectors
stationary in time.

\section{Evolution of SU(1, 1) and SU(2) non-Hermitian time-dependent systems}

The SU(1, 1) and SU(2) non-Hermitian time-dependent systems that we consider
are described by the Hamiltonian%
\begin{equation}
H(t)=2\omega(t)K_{0}+2\alpha(t)K_{-}+2\beta(t)K_{+}, \label{HH}%
\end{equation}
where $\left(  \omega(t),\alpha(t),\beta(t)\right)  $ $\in C$ are arbitrary
functions of time. $K_{0}$ is a Hermitian operator, while $K_{+}=\left(
K_{-}\right)  ^{+}$. The commutation relations between these operators are%

\begin{equation}
\left\{
\begin{array}
[c]{c}%
\left.  \left[  K_{0},K_{+}\right]  =K_{+}\right. \\
\left.  \left[  K_{0},K_{-}\right]  =-K_{-}\right. \\
\left.  \left[  K_{+},K_{-}\right]  =DK_{0}\right.
\end{array}
\right.  . \label{lie}%
\end{equation}
The Lie algebra of SU(1, 1) and SU(2) consists of the generators $K_{0}$,
$K_{-}$ and $K_{+}$ corresponding to $D=-2$ \ and $2$ in the commutation
relations (\ref{lie}), respectively.

\ With the requirement that the solution looks like a product of a
wavefunction satisfying a stationary equation multiplied by a time dependent
function, we perform the time-dependent non-unitary transformation
\begin{equation}
V(t)=\exp\left[  2\varepsilon(t)K_{0}+2\mu(t)K_{-}+2\mu^{\ast}(t)K_{+}\right]
, \label{9}%
\end{equation}
where $\varepsilon$ , $\mu$ are arbitrary real \ and complex time-dependent
parameters respectively. The group element in (\ref{9}) can be decomposed
according to \cite{kli,bar}
\begin{equation}
V(t)=e^{\vartheta_{+}(t)K_{+}}e^{\ln\vartheta_{0}(t)K_{0}}e^{\vartheta
_{-}(t)K_{-}}. \label{10}%
\end{equation}
We require here the variant (\ref{10}) of our ansatz to be able to compute the
time derivatives of $V(t)$. The time dependent coefficients read
\begin{align}
\vartheta_{0}(t)  &  =\left(  \cosh\theta-\frac{\varepsilon}{\theta}%
\sinh\theta\right)  ^{-2}\nonumber\\
\theta &  =\sqrt{\varepsilon^{2}+2D\left\vert \mu\right\vert ^{2}}\nonumber\\
\vartheta_{+}(t)  &  =\frac{2\mu^{\ast}\sinh\theta}{\theta\cosh\theta
-\varepsilon\sinh\theta}\label{11}\\
\vartheta_{-}(t)  &  =\frac{2\mu\sinh\theta}{\theta\cosh\theta-\varepsilon
\sinh\theta}\text{ }.\nonumber
\end{align}
The notation may be simplified even further by introducing some new quantities
\cite{fring2}
\begin{align}
z  &  =\frac{2\mu}{\varepsilon}=\left\vert z\right\vert e^{i\varphi
}\nonumber\\
\phi &  =\frac{\left\vert z\right\vert }{1-\frac{\varepsilon}{\theta}%
\coth\theta}\label{12}\\
\chi(t)  &  =-\frac{\cosh\theta+\frac{\epsilon}{\theta}\sinh\theta}%
{\cosh\theta-\frac{\epsilon}{\theta}\sinh\theta}\text{ }.\nonumber
\end{align}
With this adopted notation, the coefficients in (\ref{11}) simplify to%
\begin{align}
\vartheta_{\pm}  &  =-\phi e^{\mp i\varphi}\nonumber\\
\vartheta_{0}  &  =-\frac{D}{2}\phi^{2}-\chi\text{ }. \label{simp}%
\end{align}
\qquad\ 

Using the relations :%
\begin{equation}
\left\{
\begin{array}
[c]{c}%
\exp\left[  \vartheta_{-}K_{-}\right]  K_{0}\exp\left[  -\vartheta_{-}%
K_{-}\right]  =K_{0}+\vartheta_{-}K_{-}\\
\exp\left[  \vartheta_{+}K_{+}\right]  K_{0}\exp\left[  -\vartheta_{+}%
K_{+}\right]  =K_{0}-\vartheta_{+}K_{+}%
\end{array}
\right.  \label{trans1}%
\end{equation}%
\begin{equation}
\left\{
\begin{array}
[c]{c}%
\exp\left[  \ln\vartheta_{0}K_{0}\right]  K_{-}\exp\left[  -\ln\vartheta
_{0}K_{0}\right]  =\frac{K_{-}}{\vartheta_{0}}\\
\exp\left[  \vartheta_{+}K_{+}\right]  K_{-}\exp\left[  -\vartheta_{+}%
K_{+}\right]  =K_{-}+D\vartheta_{+}K_{0}-\frac{D}{2}\vartheta_{+}^{2}K_{+}%
\end{array}
\right.  \label{trans2}%
\end{equation}%
\begin{equation}
\left\{
\begin{array}
[c]{c}%
\exp\left[  \ln\vartheta_{0}K_{0}\right]  K_{+}\exp\left[  -\ln\vartheta
_{0}K_{0}\right]  =\vartheta_{0}K_{+}\\
\exp\left[  \vartheta_{-}K_{-}\right]  K_{+}\exp\left[  \vartheta_{-}%
K_{-}\right]  =K_{+}-D\vartheta_{-}K_{0}-\frac{D}{2}\vartheta_{-}^{2}K_{-}%
\end{array}
\right.  \label{trans3}%
\end{equation}
we obtain, after some algebra, the transformed Hamiltonian
\begin{equation}
\mathcal{H}(t)=2\mathcal{W}\left(  t\right)  K_{0}+2\mathcal{Q}\left(
t\right)  K_{-}+2\mathcal{Y}\left(  t\right)  K_{+}, \label{13}%
\end{equation}
where the coefficient functions are%

\begin{align}
\mathcal{W}\left(  t\right)   &  =\frac{1}{\vartheta_{0}}\left[  \omega\left(
\frac{D}{2}\vartheta_{+}\vartheta_{-}-\chi\right)  +D\left(  \vartheta
_{+}\alpha+\vartheta_{-}\beta\chi\right)  +\frac{i}{2}\left(  \dot{\vartheta
}_{0}+D\vartheta_{+}\dot{\vartheta}_{-}\right)  \right]  ,\label{W}\\
\mathcal{Q}\left(  t\right)   &  =\frac{1}{\vartheta_{0}}\left[
\omega\vartheta_{-}+\alpha-\frac{D}{2}\beta\vartheta_{-}^{2}+i\frac
{\overset{\cdot}{\vartheta_{-}}}{2}\right]  ,\text{ \ \ \ }\label{Q}\\
\mathcal{Y}\left(  t\right)   &  =\frac{1}{\vartheta_{0}}\left[  \omega
\chi\vartheta_{+}-\frac{D}{2}\alpha\vartheta_{+}^{2}+\beta\chi^{2}+\frac{i}%
{2}\left(  \vartheta_{0}\overset{\cdot}{\vartheta_{+}}-\vartheta_{+}%
\dot{\vartheta}_{0}-\frac{D}{2}\vartheta_{+}^{2}\overset{\cdot}{\vartheta_{-}%
}\right)  \right]  . \label{Y}%
\end{align}

The central idea in this procedure is to simplify the transformed Hamiltonian
$\mathcal{H}(t)$ governing the evolution of $\left\vert \phi_{n}%
(t)\right\rangle $ by cancelling the terms $K_{-}$ and $K_{+}$ and requiring
that the time evolution of a quantum system preserves the inner products
between the associated states $\langle\phi_{n}(t)\left\vert \phi
_{n}(t)\right\rangle =\langle\phi_{n}(0)\left\vert \phi_{n}(0)\right\rangle $
. This is achieved by requiring $\mathcal{Q}\left(  t\right)  =0$,
$\mathcal{Y}\left(  t\right)  =0$ and $\operatorname{Im}\mathcal{W}\left(
t\right)  =0$. These conditions impose, by using Eqs.(\ref{simp}) and after
some algebra,\ the following constraints
\begin{equation}
\overset{\cdot}{\varphi}=2\left\vert \omega\right\vert \cos\varphi_{\omega
}-2\frac{\left\vert \alpha\right\vert }{\phi}\cos(\varphi_{\alpha}%
-\varphi)+D\phi\left\vert \beta\right\vert \cos(\varphi+\varphi_{\beta}),
\label{cont1}%
\end{equation}

\begin{equation}
\overset{\cdot}{\phi}=-2\phi\left\vert \omega\right\vert \sin\varphi_{\omega
}+2\left\vert \alpha\right\vert \sin(\varphi_{\alpha}-\varphi)-D\phi
^{2}\left\vert \beta\right\vert \sin(\varphi+\varphi_{\beta}),\text{ }
\label{cont2}%
\end{equation}

\begin{equation}
\dot{\vartheta}_{0}=\frac{2\vartheta_{0}}{\phi}\left[  -2\phi\left\vert
\omega\right\vert \sin\varphi_{\omega}+\left\vert \alpha\right\vert
\sin(\varphi_{\alpha}-\varphi)+\left(  \chi-D\phi^{2}\right)  \left\vert
\beta\right\vert \sin(\varphi+\varphi_{\beta})\right]  , \label{cont3}%
\end{equation}
by which $\vartheta_{-}$, $\vartheta_{+}$ and $\vartheta_{0}$ are detemined
for given values of $\omega(t)$, $\alpha(t)$ and $\beta(t)$. It is important
to note here that when considering the time-dependent coefficient $\mu$ to be
real function instead of complex one, i.e., the polar angles $\varphi$ vanish,
the auxiliary equations (\ref{cont1})-(\ref{cont3}) that appear automatically
in this process are identical to equations (28)-(30) for Maamache et al
\cite{mus3} who used the general method of Lewis and Riesenfield to derive
them. Then the transformed Hamiltonian $\mathcal{H}(t)$ becomes%

\begin{align}
\mathcal{H}(t)  &  =2\operatorname*{Re}\left(  \mathcal{W}\left(  t\right)
\right)  K_{0}\nonumber\\
\operatorname*{Re}\left(  \mathcal{W}\left(  t\right)  \right)   &  =\left[
\left\vert \omega\right\vert \cos\varphi_{\omega}+D\phi\left\vert
\beta\right\vert \cos(\varphi+\varphi_{\beta})\right]  \text{ } \label{TH}%
\end{align}

The implication of the results is clear. The original time-dependent
quantum-mechanical problem posed through the Hamiltonian (\ref{HH}) is
completely solved if the wave function for the related transformed Hamiltonian
$\mathcal{H}(t)$ defined in Eq. (\ref{7}) is obtained. The exact solution of
the original equation (\ref{1}) can now be found by combining the above
results. We finally obtain
\begin{equation}
\left\vert \psi_{n}(t)\right\rangle =\exp\left(  i\lambda_{n}%
{\displaystyle\int\limits_{0}^{t}}
2\left[  \left\vert \omega\right\vert \cos\varphi_{\omega}+D\phi\left\vert
\beta\right\vert \cos(\varphi+\varphi_{\beta})\right]  dt^{\prime}\right)
V^{-1}(t)\left\vert \zeta_{n}\right\rangle . \label{14}%
\end{equation}

Now, weconsider the SU(1,1) case first where $D=-2$. The SU(1,1) Lie algebra
has a realization in terms of boson creation and annihilation operators
$a^{+}$ and $a$ such that%
\begin{equation}
K_{0}=\frac{1}{2}\left(  a^{+}a+\frac{1}{2}\right)  ,\text{ \ \ }K_{-}%
=\frac{1}{2}a^{2},\text{ \ \ \ \ }K_{+}=\frac{1}{2}a^{+2}. \label{15}%
\end{equation}
Then, the Hamiltonian (\ref{HH}) describes the generalized time dependent
Sawson Hamiltonian \cite{fring2}. If $\omega(t)$, $\alpha(t)$ and $\beta(t)$
are reals constant, this Hamiltonian has been studied extensively in the
literature by several authors, for instance
\cite{ahmed,swanson,jones,bagchi,musumbu,quesne,sinha,eva}. Substitution of
$D=-2$, and $\lambda_{n}=\frac{1}{2}(n+\frac{1}{2})$ into (\ref{14}) yields%

\begin{equation}
\left\vert \psi_{n}(t)\right\rangle =\exp\left(  i(n+\frac{1}{2})%
{\displaystyle\int\limits_{0}^{t}}
\left[  \left\vert \omega\right\vert \cos\varphi_{\omega}-2\phi\left\vert
\beta\right\vert \cos(\varphi+\varphi_{\beta})\right]  dt^{\prime}\right)
V^{-1}(t)\left\vert n\right\rangle , \label{16}%
\end{equation}
where $\left\vert \zeta_{n}\right\rangle =\left\vert n\right\rangle $ are the
eigenvectors of $K_{0}.$

For $D=2,$ Hamiltonian (\ref{HH})) possesses the symmetry of the dynamical
group SU(2). \ A spin in a complex time-varying magnetic field is a practical
example in this case \cite{01,02,03,04,05,06,07}. Let $K_{0}=J_{z}$
and$\ \ K_{\mp}=J_{\mp}$ . $\left\vert \zeta_{n}\right\rangle =$ $\left\vert
j,n\right\rangle $ are the eigenvectors of $J_{z}$ , i.e. $J_{z}$ $\left\vert
j,n\right\rangle =n\left\vert j,n\right\rangle $. The next step is the
calculation of the solutions (\ref{14}) which are given by \ \ \ \ %

\begin{equation}
\left\vert \psi_{n}(t)\right\rangle =\exp\left(  in%
{\displaystyle\int\limits_{0}^{t}}
\left[  \left\vert \omega\right\vert \cos\varphi_{\omega}+2\phi\left\vert
\beta\right\vert \cos(\varphi+\varphi_{\beta})\right]  dt^{\prime}\right)
V^{-1}(t)\left\vert j,n\right\rangle , \label{17}%
\end{equation}

\section{Conclusion}

We adapted the method based on a time-dependent unitary transformation of
time-dependent Hermitian Hamiltonians \cite{mus1} to solve the Schr\"{o}dinger
equation for the time-dependent non-Hermitian Hamiltonian. Starting with the
original time-dependent non-Hermitian Hamiltonian $H(t)$ and through a
non-unitary transformation $V(t)$ we derive the transformed $\mathcal{H}(t)$
as time independent Hamiltonian multiplied by a time-dependent factor. Then,
we postulate that the time evolution of a non Hermitian quantum system
preserves, not just the normalization of the quantum states, but also the
inner products between the associated states, which allows us to identify this
transformed Hamiltonian $\mathcal{H}(t)=2\operatorname*{Re}\left(
\mathcal{W}\left(  t\right)  \right)  K_{0}$ as Hermitian .\bigskip Thus, our
problem is completely solved.

Evidently, we then have presented to illustrate this theory: the SU(1, 1) and
SU(2) non-Hermitian time-dependent systems described by the Hamiltonian
(\ref{HH}) when applying the non-unitray transformation $V(t)$ we obtain the
transformed Hamiltonian $\mathcal{H}(t)$ as linear combination of $K_{0}$ and
$K_{\mp}$ .Consequently, we must disregard the prefactors of the operators
$K_{\mp}$ . To this end, we next require that the coefficients $\mathcal{Q}%
\left(  t\right)  =0$ and $\mathcal{Y}\left(  t\right)  =0$ defined in Eqs.
(\ref{Q})-(\ref{Y}). Then, by using the postulat that the inner products
between the associated states is preserved allows us to require that
$\operatorname{Im}\mathcal{W}\left(  t\right)  =0$ and to identify the
transformed Hamiltonian $\mathcal{H}(t)=2\operatorname*{Re}\left(
\mathcal{W}\left(  t\right)  \right)  K_{0}$ as Hermitian.\bigskip\ Finaly, we
also found the exact solutions of the generalized Swanson model \ and a
spinning particle in a time-varying complex magnetic field.

\end{document}